\documentclass[conference]{IEEEtran}
\usepackage{cite}
\usepackage{amsmath,amssymb,amsfonts}
\usepackage{algorithmic}
\usepackage{graphicx}
\usepackage{textcomp}
\usepackage{xcolor}
\usepackage{tabularx}
\usepackage{booktabs}
\def\BibTeX{{\rm B\kern-.05em{\sc i\kern-.025em b}\kern-.08em
    T\kern-.1667em\lower.7ex\hbox{E}\kern-.125emX}}
\ifCLASSOPTIONcompsoc
    \usepackage[caption=false, font=normalsize, labelfont=sf, textfont=sf]{subfig}
\else
\usepackage[caption=false, font=footnotesize]{subfig}
\fi
    
\newcommand{\norm}[1]{\left\lVert#1\right\rVert}

\date{}
    
\begin{document}


\title{Position-agnostic Algebraic Estimation of 6G V2X MIMO Channels via Unsupervised Learning}
\author{
\IEEEauthorblockN{Lorenzo Cazzella\IEEEauthorrefmark{1}, Dario Tagliaferri\IEEEauthorrefmark{1}, Marouan Mizmizi\IEEEauthorrefmark{1}, Matteo Matteucci\IEEEauthorrefmark{1}, \\ Damiano Badini\IEEEauthorrefmark{2}, Christian Mazzucco\IEEEauthorrefmark{2} and Umberto Spagnolini\IEEEauthorrefmark{1}}

\IEEEauthorblockA{\IEEEauthorrefmark{1}Politecnico di Milano, Milan, Italy} 
\IEEEauthorblockA{\IEEEauthorrefmark{2}Huawei Technologies Italia S.r.l., Segrate, Italy}

E-mails: \{lorenzo.cazzella, dario.tagliaferri, marouan.mizmizi, matteo.matteucci, umberto.spagnolini\}@polimi.it \\ \{damiano.badini, christian.mazzucco\}@huawei.com

}


\maketitle

\begin{abstract}
MIMO systems in the context of 6G Vehicle-to-Everything (V2X) will require an accurate channel knowledge to enable efficient communication. Standard channel estimation techniques, such as Unconstrained Maximum Likelihood (U-ML), are extremely noisy in massive MIMO settings, while structured approaches, e.g., compressed sensing, are suited to low-mobility scenarios and are sensitive to hardware impairments. We propose a novel Multi-Vehicular algebraic channel estimation method for 6G V2X based on unsupervised learning which exploits recurrent vehicle passages in typical urban settings. Multiple training sequences are clustered via K-medoids algorithm based on their \textit{algebraic similarity} to retrieve the MIMO channel eigenmodes, which can be used to improve the channel estimates. Numerical results show remarkable benefits of the proposed method in terms of Mean Squared Error (MSE) compared to standard U-ML solution (15 dB less).
\end{abstract}

\begin{IEEEkeywords} Algebraic MIMO channel estimation, 6G, V2X, Unsupervised learning, Clustering, K-medoids
\end{IEEEkeywords}

\section{Introduction}\label{sect:introduction}

Millimeter Wave (mmWave) ($30-100$ GHz) and sub-THz ($100-300$ GHz) bands definitely emerged as viable solutions for 5G and mostly 6G Vehicle-to-Everything (V2X) applications, due to the spectrum crunch at sub-6 GHz frequencies. In particular, the $24.25-52.6$ GHz band is used in 5G New Radio (NR) Frequency Range 2 (FR2), while future 6G V2X systems will exploit even larger spectrum portions in D-band ($120$ GHz)~\cite{Wymeersch2021_6G_short}. Notwithstanding, the high path loss emerging at high frequencies induces a coverage reduction and a \textit{sparse} propagation channel, characterized by few significant paths in the Space-Time (ST) domain, i.e., Angles of Arrival/Departure (AoAs/AoDs) and delays~\cite{6834753}. To compensate for the path loss, Multiple-Input Multiple-Output (MIMO) antenna systems at both Transmitter (Tx) and Receiver (Rx) are required to increase antenna gain by beamforming strategies~\cite{7244171}.

In massive MIMO systems, the channel knowledge becomes essential for designing the correct beamforming directions at both Tx and Rx sides. In Orthogonal Frequency Division Multiplexing/Multiple Access (OFDM/OFDMA) systems, such as the 5G NR FR2 radio interface, standard approaches are based on an Unconstrained Maximum Likelihood (U-ML) channel estimate at each training block from known pilot sequences, which however is extremely noisy in low Signal-to-Noise Ratio (SNR) conditions. Its performance can be improved with constrained approaches reducing the number of unknowns operating on multiple channel realizations, and this can be obtained either via a structured approach, such as Compressed Sensing (CS) \cite{5454399}, or via an unstructured approach, recognizing the algebraic Low-Rank (LR) nature of the MIMO channel matrix \cite{Nicoli2003}. Although providing remarkable performance, CS is sensitive to array system calibrations and parameters initialization \cite{Cerutti2020}, being currently too complex for rapidly time-varying V2X channels.

LR channel estimation methods, originally proposed for sub-6 GHz systems \cite{Nicoli2003} and then for mmWave \cite{Cerutti2020,Nicoli2020}, operate on ensembles of training sequences from a single (or multiple) User Equipment (UE) to a fixed Base Station (BS), leveraging on the algebraic channel sparsity and on the stationarity of the ST eigenmodes (ST subspace) in time \cite{Cerutti2020} or space \cite{Nicoli2020}, i.e., the invariance of AoAs/AoDs and delays over multiple channel realizations. The improved LR channel estimate is retrieved by a modal filtering of the received training sequence onto the ST propagation subspace. LR achieves the same advantages of CS but with an inherent robustness against hardware impairments \cite{Cerutti2020}. Nevertheless, the main drawback of the aforementioned LR implementation is that it requires either a large number of consecutive transmissions (not suited to V2X) or the knowledge of UEs position at the BS (increased control signaling). 

In this work, we take advantage of the Multi-Vehicular (MV) approach proposed in \cite{Nicoli2020} to design a novel \textit{position-agnostic} clustering-based MV-LR channel estimation method suited for V2X, where the channel eigenmodes are retrieved from the set of received training sequences collected from the \textit{recurrent vehicle passages} in a limited urban area (radio cell), where the road constraints induce recurrences in the ST MIMO channel subspace. 

Instead of requiring the explicit knowledge of UEs positions, by leveraging the works in \cite{jung2018pilotless,azizipour2019channel}, we frame the problem of obtaining the ensemble of received training sequences for LR as a K-medoids high-dimensional clustering problem in the ST subspace of the MIMO channel, where different received training sequences are grouped based on their algebraic similarity. Numerical simulations with ray-tracing channel data and realistic vehicle trajectories show the effectiveness of the proposed method, highlighting a Mean Squared Error (MSE) gain in the order of $\approx 15$ dB compared to U-ML channel estimation, attaining the theoretical bound.

\textit{Notation}: Bold upper- and lower-case letters describe matrices and column vectors. $(\cdot)^{\mathrm{T}}$, $(\cdot)^{\mathrm{H}}$, $(\cdot)^{*}$, $\norm{\cdot}$, and $\mathrm{vec}(\cdot)$ denote, respectively, transpose, conjugate-transpose, conjugate, Frobenius norm, and vectorization by columns of a matrix. $\mathrm{tr}\left(\cdot\right)$, $\mathrm{rank}\left(\cdot\right)$, $\mathrm{eig}_{r}(\cdot)$, $\mathrm{span}(\cdot)$, extract the trace, the rank, the $r$ eigenvectors and the subspace spanned by the columns of a matrix. $\otimes$, $\diamond$ and $\odot$ denote the Kronecker, the Kathri-Rao (column-wise) and the element-wise product between two matrices. $\mathbf{A}^{\dagger}$ is the Moore-Penrose pseudo-inverse of $\mathbf{A}$. $\mathrm{diag}(\cdot)$ denotes either a diagonal matrix or the extraction of the diagonal of a matrix. $\mathbf{a}\sim\mathcal{CN}(\boldsymbol{\mu},\mathbf{C})$ denotes a multi-variate complex Gaussian random variable $\mathbf{a}$ with mean $\boldsymbol{\mu}$ and covariance $\mathbf{C}$. $\mathbb{E}[\cdot]$ is the expectation operator, while $\mathbb{R}$ and $\mathbb{C}$ stand for the set of real and complex numbers, respectively. Finally, $\delta_{n}$ is the Kronecker delta.

\section{System and Channel Model}\label{sect:system&channelmodel}

We consider a single-user, multi-carrier MIMO uplink system tailored to a V2I scenario, in which the Tx and the Rx are equipped with $N_T$ and $N_R$ antennas, respectively. 
We assume the available spectrum divided in $N_K$ subcarriers. 
At the receiving antennas, after the time and frequency synchronization and cyclic prefix removal, the Rx signal is:
\begin{equation}\label{eq:Rxsignal}
    \mathbf{y}(t) = \mathbf{H}(t) * \mathbf{x}(t) + \mathbf{n}(t),
\end{equation}
where symbol $*$ denotes the matrix convolution between the transmitted signal $\mathbf{x}(t)\in\mathbb{C}^{N_T \times 1}$ and the frequency-selective MIMO channel $\mathbf{H}(t)\in\mathbb{C}^{N_R \times N_T}$. Vector $\mathbf{n}(t)\in\mathbb{C}^{N_R \times 1}$ denotes the additive Gaussian noise corrupting the received signal. Sampling \eqref{eq:Rxsignal} at time $t=wT$, with $T=1/B$ being the sampling time ($B$ the bandwidth) and performing the $N_K$-point Discrete Fourier Transform (DFT) we obtain:
\begin{equation}\label{eq:Rxsignal_frequency}
    \mathbf{y}[k] = \mathbf{H}[k]\, \mathbf{x}[k] + \mathbf{n}[k],
\end{equation}
where $\mathbf{H}[k]$ is the MIMO channel per-subcarrier. For channel estimation purposes, the Tx signal is a known training sequence assumed to be random (but known at receiver) and uncorrelated in space and frequency as $\mathbb{E}\left[\mathbf{x}[k] \mathbf{x}[\ell]^{\mathrm{H}}\right] = \sigma^2_x \mathbf{I}_{N_T} \delta_{k-\ell}$. The noise $\mathbf{n}[k]$ is instead modelled as white in time/frequency, but generally colored in space, to account for directional interference, as
$\mathbb{E}\left[\mathbf{n}[k] \mathbf{n}[\ell]^{\mathrm{H}}\right] = \mathbf{Q}_n \delta_{k-\ell}$. The average SNR per subcarrier is
$\mathrm{SNR} = \frac{1}{N_K} \sum_k \mathbb{E}[\norm{\mathbf{H}[k] \mathbf{x}[k]}^2]/\mathbb{E}[\norm{\mathbf{n}[k]}^2]$.

\subsection{MIMO Channel Model}

The discrete-time wideband (frequency-selective) MIMO channel between Tx and Rx is routinely modeled as the sum of $P$ paths as \cite{6834753}
\begin{equation}\label{eq:channel_matrix_tap_time} 
\begin{split}
        \mathbf{H}[w] & = \sum_{p=1}^{P}\alpha_p\,\mathbf{a}_R(\boldsymbol{\vartheta}_p)\mathbf{a}_T^{\mathrm{T}}(\boldsymbol{\psi}_p)g\left[(w-1)T-\tau_p\right] = \\
        & = \mathbf{A}_R\left(\boldsymbol{\vartheta}\right)
        \boldsymbol{\Lambda}[w]\mathbf{A}_T^{\mathrm{T}}\left(\boldsymbol{\psi}\right),
\end{split}
\end{equation}
where: 
\begin{itemize}

    \item $\mathbf{H}[w]\in\mathbb{C}^{N_R\times N_T}$, $w=1,\dots,W$, is the $w$-th tap (out of $W$) of the discrete-time channel matrix;
    
    \item $\alpha_p \sim \mathcal{CN}\left(0,\Omega_p\right)$ is the complex gain of the $p$-th path. The paths' amplitudes $\boldsymbol{\alpha} = [\alpha_1,\dots,\alpha_P]^\mathrm{T}$ are assumed to the Wide-Sense Stationary Uncorrelated Scattering (WSSUS) model, i.e., $\mathbb{E}\left[ \boldsymbol{\alpha}_n \boldsymbol{\alpha}^{\mathrm{H}}_{n+m}\right] = \boldsymbol{\Omega}\,\delta_{n-m}$, in which $\boldsymbol{\Omega} = \mathrm{diag}\left(\Omega_1, \dots,\Omega_P\right)$ contains the paths' powers and $n$, $m$ are two channel realizations in time (different fading blocks) or space (different locations);
    
    \item $\mathbf{a}_T(\boldsymbol{\psi}_p)\in\mathbb{C}^{N_T \times 1}$ and $\mathbf{a}_R(\boldsymbol{\vartheta}_p)\in\mathbb{C}^{N_R \times 1}$ are the Tx and Rx array response vectors to the $p$-th path, respectively, function of the DoDs $\boldsymbol{\psi}_p$ and the DoAs $\boldsymbol{\vartheta}_p$;
    
    \item $g\left[(w-1)T-\tau_p\right]$ denotes the $w$-th sample of the cascade response of the Tx and the Rx pulse shaping filters (PSF), delayed by $\tau_p$ ($p$-th path delay).
    
\end{itemize}

In \eqref{eq:channel_matrix_tap_time}, matrices $\mathbf{A}_T\left(\boldsymbol{\psi}\right) = \left[\mathbf{a}_T(\boldsymbol{\psi}_1),\dots,\mathbf{a}_T(\boldsymbol{\psi}_P)\right] \in \mathbb{C}^{N_T\times P}$ and
$\mathbf{A}_R\left(\boldsymbol{\vartheta}\right) = \left[\mathbf{a}_R(\boldsymbol{\vartheta}_1),\dots,\mathbf{a}_R(\boldsymbol{\vartheta}_P)\right] \in \mathbb{C}^{N_R\times P}$  identify the Tx and Rx frequency-independent beam spaces, respectively, while $\boldsymbol{\Lambda}[w] =  \mathrm{diag}(\alpha_1\,g[(w-1)T-\tau_1],\dots,\alpha_P\,g[(w-1)T-\tau_P]) \in \mathbb{C}^{P\times P}$ is a diagonal matrix collecting all the channel amplitudes scaled by the $w$-th tap of the PSF. Matrices $\mathbf{A}_T\left(\boldsymbol{\psi}\right)$ and $\mathbf{A}_R\left(\boldsymbol{\vartheta}\right)$ define the Tx and Rx diversity orders of channel $\mathbf{H}[w]$ as 
\begin{align}
    r_{\mathrm{S}}^{\mathrm{TX}} & = \mathrm{rank}(\mathbf{A}_T\left(\boldsymbol{\psi}\right)) \leq \mathrm{min}\left(N_T, P\right),\label{eq:TX_diversity_order}\\
    r_{\mathrm{S}}^{\mathrm{RX}} & = \mathrm{rank}(\mathbf{A}_R\left(\boldsymbol{\vartheta}\right))  \leq \mathrm{min}\left(N_R, P\right),\label{eq:RX_diversity_order}
\end{align}
i.e., the number of resolvable spatial paths given the number of Tx and Rx antennas, respectively. 
The channel for frequency $k$ in \eqref{eq:channel_matrix_tap_time} can be obtained from $\mathbf{H}[w]$ with a Discrete Fourier Transform (DFT).
%
%
By rearranging channel \eqref{eq:channel_matrix_tap_time}, we can isolate the temporal (delays) features of the channel as:
\begin{equation}\label{eq:channel_ST_matrix}
    \boldsymbol{\mathcal{H}} =\boldsymbol{\mathcal{A}}\left(\boldsymbol{\vartheta},\boldsymbol{\psi}\right) \mathbf{D} \,\mathbf{G}^{\mathrm{T}}(\boldsymbol{\tau}),
\end{equation}
where: \textit{(i)} $\boldsymbol{\mathcal{H}} = [\mathrm{vec}(\mathbf{H}[1]),\dots,\mathrm{vec}(\mathbf{H}[W])]$; \textit{(ii)} $\boldsymbol{\mathcal{A}}\left(\boldsymbol{\vartheta},\boldsymbol{\psi}\right) = \mathbf{A}_T(\boldsymbol{\psi}) \diamond \mathbf{A}_R(\boldsymbol{\vartheta})\in\mathbb{C}^{N_T N_R \times P}$ span the \textit{joint} Tx and Rx beam space; \textit{(iii)} $\mathbf{D} = \mathrm{diag}(\alpha_1,\dots,\alpha_P)$ and \textit{(iv)} $\mathbf{G}\left(\boldsymbol{\tau}\right) = \left[\mathbf{g}(\tau_1),\dots,\mathbf{g}(\tau_P)\right]$ embed the temporal features $\tau$. Vector $\mathbf{g}\left(\tau_p\right) \in\mathbb{R}^{W \times 1} = \left[g\left[-\tau_p\right],\dots,g\left[(W-1)T-\tau_p\right]\right]^{\mathrm{T}}$ collects PSF samples delayed by $\tau_p$. With this channel formulation, the temporal diversity order is:
\begin{equation}\label{eq:T_diversity_order}
     r_{\mathrm{T}} = \mathrm{rank}(\mathbf{G}\left(\boldsymbol{\tau}\right))  \leq \mathrm{min}\left(W, P\right).
\end{equation}

%
%
%
%

\section{MV-LR MIMO Channel Estimation}

To overcome the limitations of U-ML channel estimation techniques, we adapt here the LR method to high-mobility V2X systems by exploiting the MV concept proposed in \cite{Nicoli2020}. The BS estimates the ST eigenmodes of channel $\widetilde{\mathbf{H}}[w]$ from the ensemble of $L$ received training sequences $\{\mathbf{y}_{\ell}[k]\}_{\ell=1}^{\ell=L}$, sharing the same ST propagation subspace, collected from \textit{recurrent vehicle passages} over the same geographical location. The underlying idea of the proposed MV-LR is that, in a quasi-static propagation environment, different vehicles (with the same antenna equipment) passing on the same location in space with only slightly different trajectories, as commonly happens in urban traffic scenarios, experience the same AoDs/AoAs and delays in communicating with the BS, and different fading amplitudes. In this context, the training sequences $\{\mathbf{y}_{\ell}[k]\}_{\ell=1}^{\ell=L}$ share the same propagation subspace. There are two possible implementations of the method, based on the available degree of cooperation between the UEs and the infrastructure (BS). In both cases, the notable advantage of MV-LR is the possibility, for the BS, to store the ST eigenmodes list, in order to avoid repeating the training procedure for each vehicle, minimizing the computations.

\textbf{Position-aware approach}: The ST eigenmodes of the MIMO channel are explicitly associated to the physical position in the cell of the UE. The BS collects the $L$ received training sequences $\{\mathbf{y}_{\ell}[k]\}_{\ell=1}^{\ell=L}$ for each location in the cell by relating them with the estimated physical UEs positions, obtained through either a suitable signaling or other localization techniques. The UEs are requested to cooperate with the infrastructure to build the database of ST channel eigenmodes, and the LR estimation performance depends on the positioning accuracy, which can be in the order of few meters in urban scenarios. 

\textbf{Position-agnostic approach}: The ST eigenmodes of the MIMO channel are \textit{not} related to a given physical UE position but rather are \textit{subspace-dependent}. A huge dataset of $N$ received training sequences $\{\mathbf{y}_{\ell}[k]\}_{i=1}^{i=N}$, $N \gg L$, not explicitly related to physical positions, is clustered at the BS with an unsupervised learning approach to \textit{automatically} devise the algebraic similarity (subspace similarity) in the dataset. The cooperation between UEs and BS is minimal (exchange of training sequences, already in place for communication), and the performance of the system depends on the number $K$ of chosen clusters, on the dataset (cardinality, data diversity), and on the selected similarity metric.

In the following, we outline the algebraic background for the LR channel estimation from $L$ different training sequences $\{\mathbf{y}_{\ell}[w]\}_{\ell=1}^{\ell=L}$, assumed to share the same ST propagation subspace. More details can be found in \cite{Nicoli2003}. 
The LR-estimated channel is retrieved through the application of a \textit{training sequence-specific} matrix $\mathbf{T}_{\ell}$ and an \textit{ensemble-specific} matrix $\boldsymbol{\Pi}(L)$ on single received training signal $\mathbf{y}_{\ell} = \left[\mathbf{y}^{\mathrm{T}}_{\ell}[1],\dots,\mathbf{y}^{\mathrm{T}}_{\ell}[N_K]\right]^{\mathrm{T}}\in\mathbb{C}^{N_K N_R \times 1}$ as:
\begin{equation}\label{eq:LR}
    \widehat{\mathbf{h}}_{\ell} = \boldsymbol{\Pi}(L)\,\mathbf{T}_{\ell} \,\mathbf{y}_{\ell} = \boldsymbol{\Pi}(L)\,\overline{\mathbf{y}}_{\ell},
\end{equation}
where: \textit{(i)} $\widehat{\mathbf{h}}_{\ell}\in\mathbb{C}^{W N_R N_T \times 1}$ is the LR-estimated channel vector, that can be rearranged to obtain either $\widehat{\mathbf{H}}_{\ell}[w]$ or $\widehat{\mathbf{H}}_{\ell}[k]$ and \textit{(ii)} $\overline{\mathbf{y}}_{\ell}=\mathbf{T}_{\ell} \,\mathbf{y}_{\ell}\in\mathbb{C}^{W N_R N_T \times 1}$ is the pre-processed signal by matrix $\mathbf{T}_{\ell}$. A notable example of pre-processing is the U-ML channel estimation, here adopted, and detailed in \cite{Cerutti2020}.

Based on the LR constraints in \eqref{eq:TX_diversity_order}, \eqref{eq:RX_diversity_order} and \eqref{eq:T_diversity_order}, the ensemble-specific matrix in \eqref{eq:LR} is designed as \cite{Nicoli2003}:
\begin{equation}\label{eq:DST_projector}
    \boldsymbol{\Pi}(L) = \widehat{\mathbf{C}}^{\frac{\mathrm{H}}{2}}\, \widehat{\boldsymbol{\Pi}}\, \widehat{\mathbf{C}}^{-\frac{\mathrm{H}}{2}},
\end{equation}
where \textit{(i)} $\widehat{\mathbf{C}}\approx (1/\sigma^2_x) \,(\mathbf{I}_{W} \otimes \mathbf{I}_{N_T} \otimes \mathbf{Q}^{\mathrm{T}}_n)$ is the estimated covariance matrix of $\overline{\mathbf{y}}_\ell$, needed to handle spatial/temporal noise correlations (e.g., interfering users) and \textit{(ii)} $\widehat{\boldsymbol{\Pi}}=\widehat{\mathbf{U}}^*_{\mathrm{T}}\widehat{\mathbf{U}}^{\mathrm{T}}_{\mathrm{T}}\otimes \widehat{\mathbf{U}}^{\mathrm{Tx},*}_{\mathrm{S}} \widehat{\mathbf{U}}^{\mathrm{Tx,T}}_{\mathrm{S}} \otimes \widehat{\mathbf{U}}^{\mathrm{Rx}}_{\mathrm{S}} \widehat{\mathbf{U}}^{\mathrm{Rx,H}}_{\mathrm{S}}$ is the projection matrix onto the ST propagation subspace associated to the separable basis $\widehat{\mathbf{U}}=\widehat{\mathbf{U}}^*_{\mathrm{T}}\otimes \widehat{\mathbf{U}}^{\mathrm{Tx,*}}_{\mathrm{S}} \otimes \widehat{\mathbf{U}}^{\mathrm{Rx}}_{\mathrm{S}} $. Orthonormal bases $\widehat{\mathbf{U}}_{\mathrm{T}}\in\mathbb{C}^{W \times r_{\mathrm{T}}}$, $\widehat{\mathbf{U}}^{\mathrm{Tx}}_{\mathrm{S}}\in\mathbb{C}^{N_T \times r^{\mathrm{Tx}}_{\mathrm{S}}}$ and  $\widehat{\mathbf{U}}^{\mathrm{Rx}}_{\mathrm{S}}\in\mathbb{C}^{N_T \times r^{\mathrm{Rx}}_{\mathrm{S}}}$ are retrieved as the $r^{\mathrm{Tx}}_{\mathrm{S}}$, $r^{\mathrm{Rx}}_{\mathrm{S}}$ and $r_{\mathrm{T}}$ leading eigenvectors of the following sample correlation matrices over $L$ received training sequences:

\begin{align}
    \widehat{\mathbf{R}}^{\mathrm{Tx}}_{\mathrm{S}} & = \frac{1}{L}\sum_{\ell=1}^{L}\sum_{w=1}^{W} \overline{\overline{\mathbf{Y}}}_{\ell}[w]\, \overline{\overline{\mathbf{Y}}}^{\mathrm{H}}_{\ell}[w], \label{eq:R_S_TX_sample}\\
    \widehat{\mathbf{R}}^{\mathrm{Rx}}_{\mathrm{S}} & = \frac{1}{L}\sum_{\ell=1}^{L}\sum_{w=1}^{W} \overline{\overline{\mathbf{Y}}}^{\mathrm{H}}_{\ell}[w]\, \overline{\overline{\mathbf{Y}}}_{\ell}[w],\label{eq:R_S_RX_sample}\\
    \widehat{\mathbf{R}}_{\mathrm{T}} & = \frac{1}{L}\sum_{\ell=1}^{L} \overline{\overline{\boldsymbol{\mathcal{Y}}}}_{\ell}^{\mathrm{H}}\, \overline{\overline{\boldsymbol{\mathcal{Y}}}}_{\ell},\label{eq:R_T_sample}
\end{align}
where $\overline{\overline{\mathbf{Y}}}_{\ell}[w]\in\mathbb{C}^{N_T\times N_R}$ and $\overline{\overline{\boldsymbol{\mathcal{Y}}}}_{\ell}\in\mathbb{C}^{N_T N_R \times W}$ are suitable rearrangements of the \textit{whitened} sequence $\overline{\overline{\mathbf{y}}}_{\ell} = \widehat{\mathbf{C}}^{-\frac{\mathrm{H}}{2}} \overline{\mathbf{y}}_{\ell}\in\mathbb{C}^{W N_T N_R \times 1}$. 

The LR performance is proportional to the \textit{unstructured sparsity degree} of the channel. 
It can be demonstrated that, if at least one of the following conditions holds:
\begin{align}\label{eq:}
  r^{\mathrm{Tx}}_{\mathrm{S}} < N_T, \;\;\;\;
  r^{\mathrm{Rx}}_{\mathrm{S}} < N_R, \;\;\;\;
  r_{\mathrm{T}} < W,
\end{align}
the LR method outperforms the U-ML one. Asymptotically ($L\rightarrow \infty$), the estimated subspaces converge to: 
\begin{align}
    &\mathrm{span}(\widehat{\mathbf{U}}^{\mathrm{Tx}}_{\mathrm{S}}) \rightarrow \mathrm{span}(\mathbf{A}_{T}(\boldsymbol{\psi})),\\
    &\mathrm{span}(\widehat{\mathbf{U}}^{\mathrm{Rx}}_{\mathrm{S}}) \rightarrow \mathrm{span}(\mathbf{Q}^{-\frac{\mathrm{H}}{2}}_n \mathbf{A}_R(\boldsymbol{\vartheta})),\\
    &\mathrm{span}(\widehat{\mathbf{U}}_{\mathrm{T}}) \rightarrow \mathrm{span}(\mathbf{G}(\boldsymbol{\tau})),
\end{align}
and the LR attains the maximum performance.
The value of $L$ for the asymptotic convergence depends on the size of the correlation matrices in \eqref{eq:R_S_TX_sample}, \eqref{eq:R_S_RX_sample} and \eqref{eq:R_T_sample} as well as on the SNR. For the MIMO settings and bandwidths considered in Section \ref{sect:results}, $L\approx 100$ guarantees the convergence.

\section{Clustering-based MV-LR Channel Estimation}\label{sec:method}
In this section, we describe the clustering algorithm used for the position-agnostic MV-LR implementation in V2X urban settings. Let us consider a large number, $N$, of received training sequences $\{\overline{\overline{\mathbf{y}}}_{i}\}_{i=1}^{i=N}$, collected at the BS over the whole radio cell, already pre-processed by matrix $\mathbf{T}_{i}$ and whitened. We aim at clustering them in order to \textit{(i)} identify few representative received training sequences with markedly different ST features, allowing to define a finite set of $K$ few comprehensive ST patterns (\textit{clusters}) easy to discriminate in a noisy setting; \textit{(ii)} compute the LR orthonormal sets $\widehat{\mathbf{U}}_{\mathrm{T}}$, $\widehat{\mathbf{U}}^{\mathrm{Tx}}_{\mathrm{S}}$ and  $\widehat{\mathbf{U}}^{\mathrm{Rx}}_{\mathrm{S}}$ for each cluster to efficiently apply LR estimation.

The proposed goals can be modelled in the framework of the $K$-medoids problem. With respect to the well-known K-means algorithm, K-medoids does not require the computation of a mean---which is meaningless for received training sequences belonging to different locations in space---, and it is more resilient to outliers and noise.

Given a set of data points $X = \{x_j\}\quad j = 1, \cdots, N$, K-medoids clustering aims at selecting $K$ elements $m_i$---called medoids---among them such that the sum of dissimilarities
\begin{equation}
    D = \sum_{k = 1}^K \sum_{x_j \in C_k} d(x_j, m_k)
\end{equation}
is minimized, where $C_k$ is the cluster represented by medoid $m_k$, and $d$ is an arbitrary dissimilarity measure between two data points. A medoid $m_k$ minimizes the intra-cluster sum of dissimilarities:
\begin{equation}
    m_k = \underset{x_j \in C_k}{\mathrm{argmin}} \sum_{x_t \in C_i} d(x_j, x_t).
\end{equation}
After a random initialization, the clusters are defined by assigning, according to the utilized dissimilarity measure, each dataset point to the nearest medoid, which can be considered a representative element of the cluster. In this work, to solve the K-medoids problem, the Partitioning Around Medoids (PAM) \cite{kaufman1990partitioning} algorithm has been used.


For grouping ST-similar received training sequences, we take advantage the subspace correlation index proposed in \cite{bosisio2006enhanced}, deriving the following similarity metric:
\begin{equation}\label{eq:similarity_function}
    \eta_{i, j} = \frac{\mathrm{tr}[\mathbf{R}_i \mathbf{R}_j^H]}{\sqrt{\mathrm{tr}[\mathbf{R}_i \mathbf{R}_i^H]\mathrm{tr}[\mathbf{R}_j \mathbf{R}_j^H]}} = 1 - d_{i, j}
\end{equation}
for $ i,j \in {1, \cdots, N }$, where $\mathbf{R}_i =\overline{\overline{\mathbf{y}}}_{i} \overline{\overline{\mathbf{y}}}^{\mathrm{H}}_{i}$, and $\eta_{i, j} \in [0, 1]$. The dissimilarity measure $d_{i, j}$ is able to capture the \textit{distance} of two received sequences $\overline{\overline{\mathbf{y}}}_{i}$, $\overline{\overline{\mathbf{y}}}_{j}$ in the ST domain, as shown in Section \ref{sect:results}. 



The proposed method can be summarized by the following steps:
\begin{enumerate}
    \item Collection at the BS of training sequences $\{\overline{\overline{\mathbf{y}}}_{i}\}_{i=1}^{i=N}$, transmitted by UEs crossing the radio cell.
    \item Clustering of the collected training sequences within the ST domain into $K$ clusters by means of the PAM algorithm, using the dissimilarity metric $d_{i, j}$ derived from \eqref{eq:similarity_function}.
    \item Computation of the MV-LR ST orthonormal bases $\{\widehat{\mathbf{U}}_{\mathrm{T}}$, $\widehat{\mathbf{U}}^{\mathrm{Tx}}_{\mathrm{S}}$,   $\widehat{\mathbf{U}}^{\mathrm{Rx}}_{\mathrm{S}}\}_{k}$, $k = 1, \cdots, K$, by using the corresponding clustered received training sequences.
    \item Filtering of the $\ell$-th new received sequence $\overline{\overline{\mathbf{y}}}_\ell$ by using the set of LR orthonormal bases corresponding to medoid $m_k,\; k = 1,\cdots, K$, nearest to $\overline{\overline{\mathbf{y}}}_\ell$ with respect to dissimilarity $d_{i, j}$.
\end{enumerate}


Since the convergence of the MV-LR algorithm is affected by the number of available received training sequences per cluster, we adopt the silhouette method \cite{rousseeuw1987silhouettes} to determine the clustering quality and to select a suitable number of clusters $K$, searching for: \textit{(i)} an even distribution of the training points among clusters to ensure the convergence of the MV-LR algorithm for each of them, \textit{(ii)} a high intra-cluster cohesion, and \textit{(iii)} a low inter-cluster similarity.

The time complexity of the PAM algorithm scales $\propto N^2$, which is affordable for the considered dataset (see Section \ref{sect:results}), but still inherently limited for very large datasets, requiring more efficient algorithms for application in practical systems. A valid alternative is CLARA (Clustering for Large Applications) \cite{KAUFMAN1986425}, which runs PAM multiple times on small subsamples of the original dataset.

\begin{table}[t!]
\centering
\caption{Simulation parameters}
\begin{tabular}{l|c|c}
\toprule
\textbf{Simulation parameter} & \textbf{Symbol} & \textbf{Value}\\
\hline
Carrier frequency & $f_0$ & $28$ GHz\\
Bandwidth & $B$ & $1$, $50$ MHz\\
BS height from the ground & - & $6$ m\\
Number of BS antennas & $N_R$ & $64$ ($8 \times 8$) \\
Number of UE antennas & $N_T$ & $16$ ($4 \times 4$)\\
Training dataset size & $N$ & $5000$ samples\\
Number of clusters & $K$ & $7$, $8$\\
Signal to Noise Ratio & SNR & $0$ dB \\ 
\bottomrule
\end{tabular}
\label{tab:simulation_parameters}
\end{table}

\section{Numerical results}\label{sect:results}

\begin{figure}
    \centering
    \includegraphics[width=0.7\columnwidth]{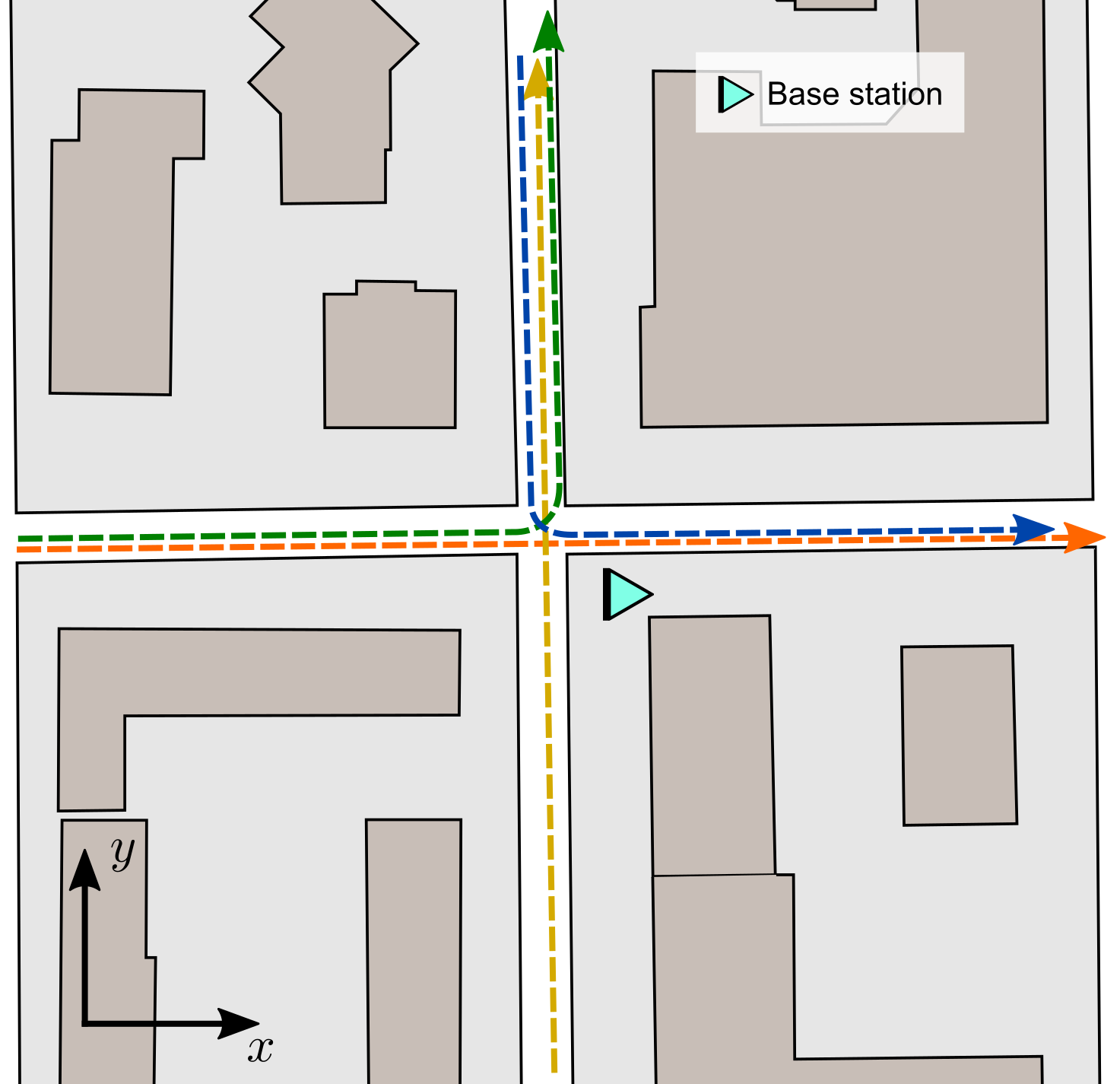}
    \caption{Selected urban scenario and representation of the considered vehicular trajectories.}
    \label{fig:map_with_trajectories}
\end{figure}

We analyse the performance of the proposed channel estimation method in the communication scenario (radio cell) depicted in Fig. \ref{fig:map_with_trajectories}. The BS, located at a height of $6$ m from the ground, is equipped with planar $64$ antennas ($8 \times 8$), while each UE with planar $16$ antennas ($4 \times 4$). We select $28$ GHz as the carrier frequency (compliant to 5G NR FR2) and two communication bandwidths: \textit{(i)} $B=1$ MHz, for which the MIMO channel is frequency-flat ($W=1$), and therefore the clustering is performed over the spatial subspaces only, ruled by the number of UE and BS antennas; \textit{(ii)} $B=50$ MHz, for which the MIMO channel is frequency-selective ($W=7$ taps), and the K-medoids is applied to the ST channel subspace. The set of simulation parameters is reported in Table \ref{tab:simulation_parameters}.

\begin{figure*}
    \centering
    \subfloat[\label{1a}]{%
       \includegraphics[width=0.4\linewidth]{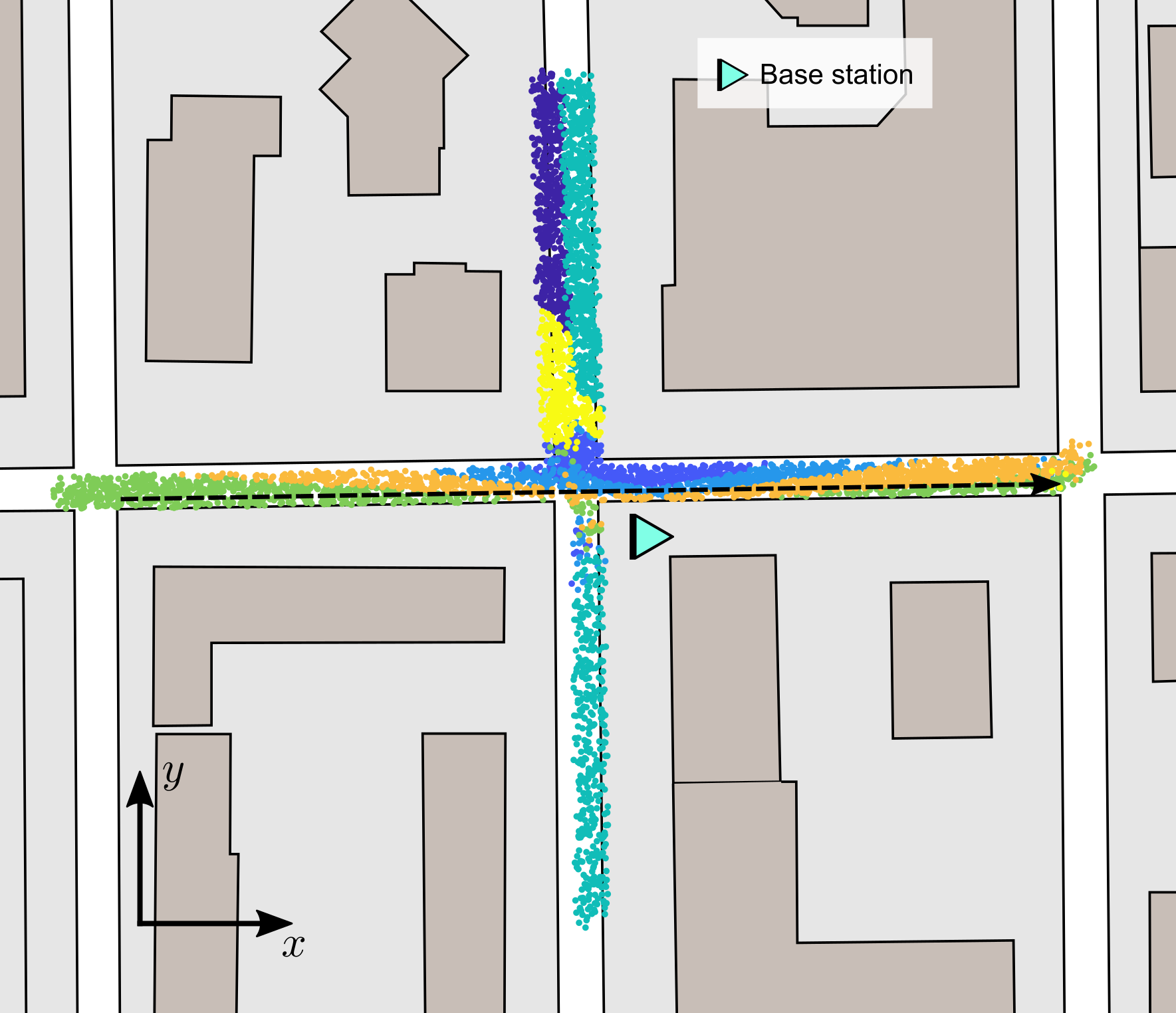}}
    \hfill
    \vrule
    \hfill
    \subfloat[\label{2a}]{%
       \includegraphics[width=0.4\linewidth]{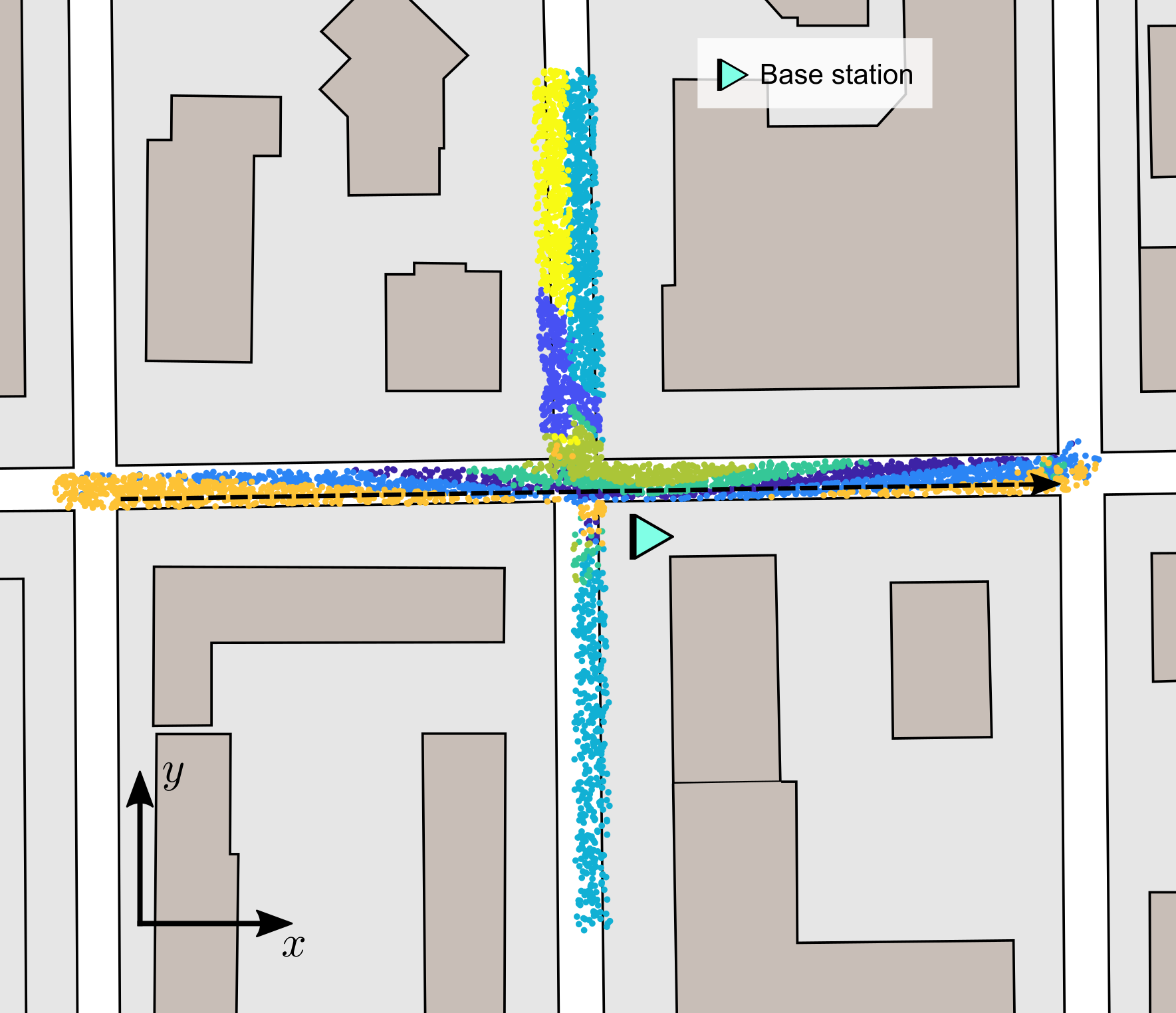}}
       
    \subfloat[\label{1c}]{%
       \includegraphics[width=0.42\linewidth]{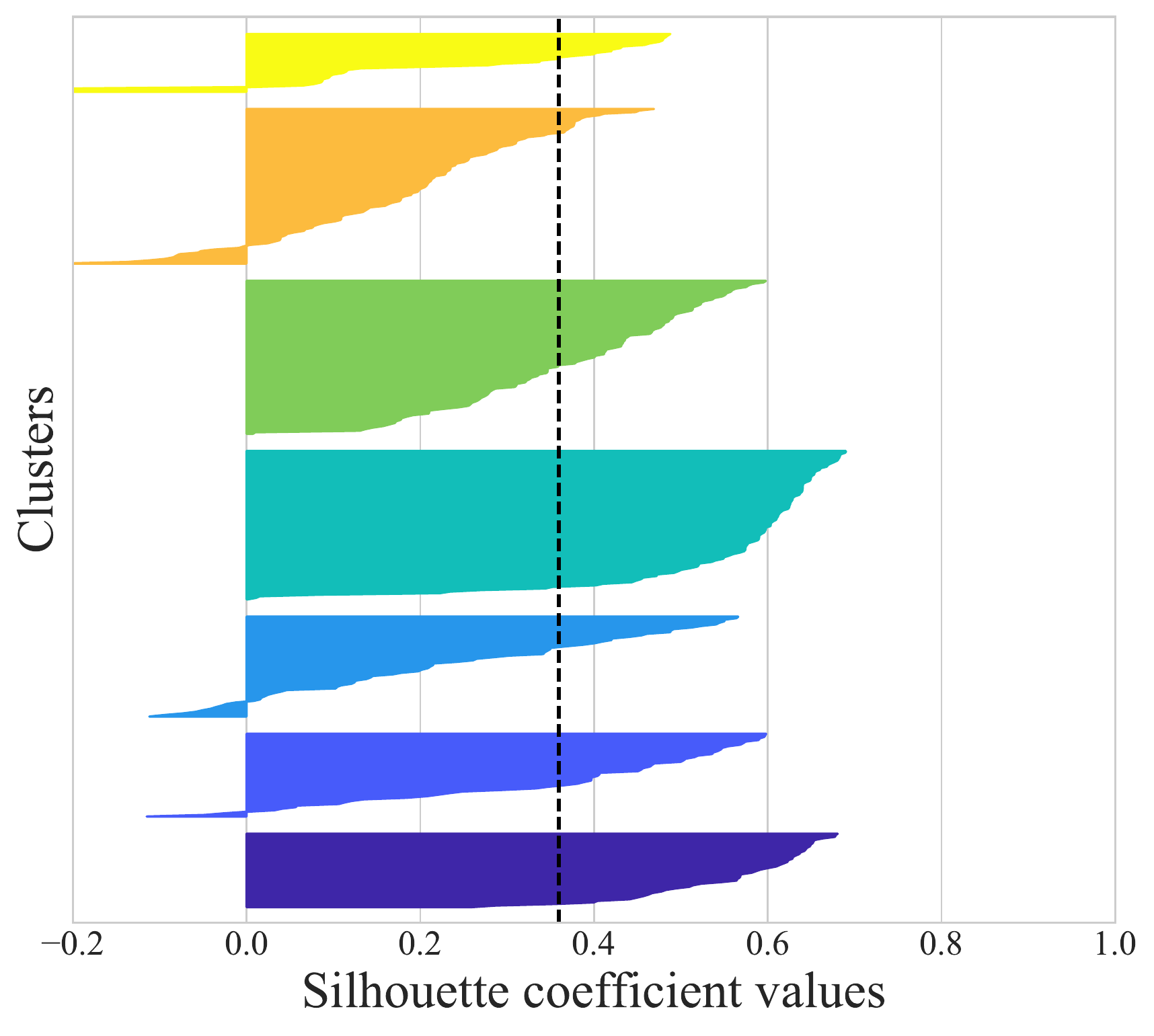}}
    \hfill
    \vrule
    \hfill
    \subfloat[\label{2c}]{%
       \includegraphics[width=0.42\linewidth]{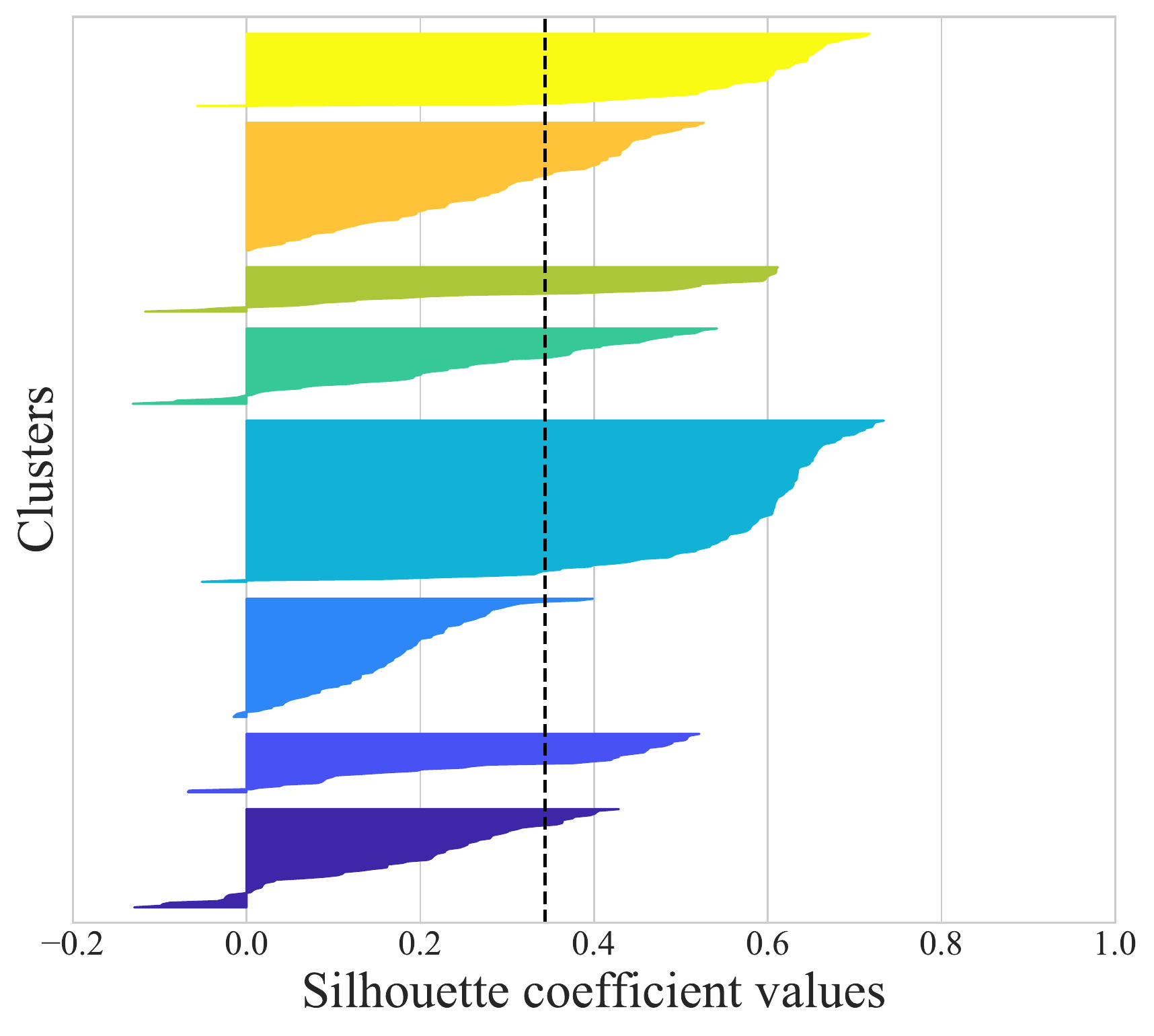}}
       
    \subfloat[\label{1b}]{%
       \includegraphics[width=0.42\linewidth]{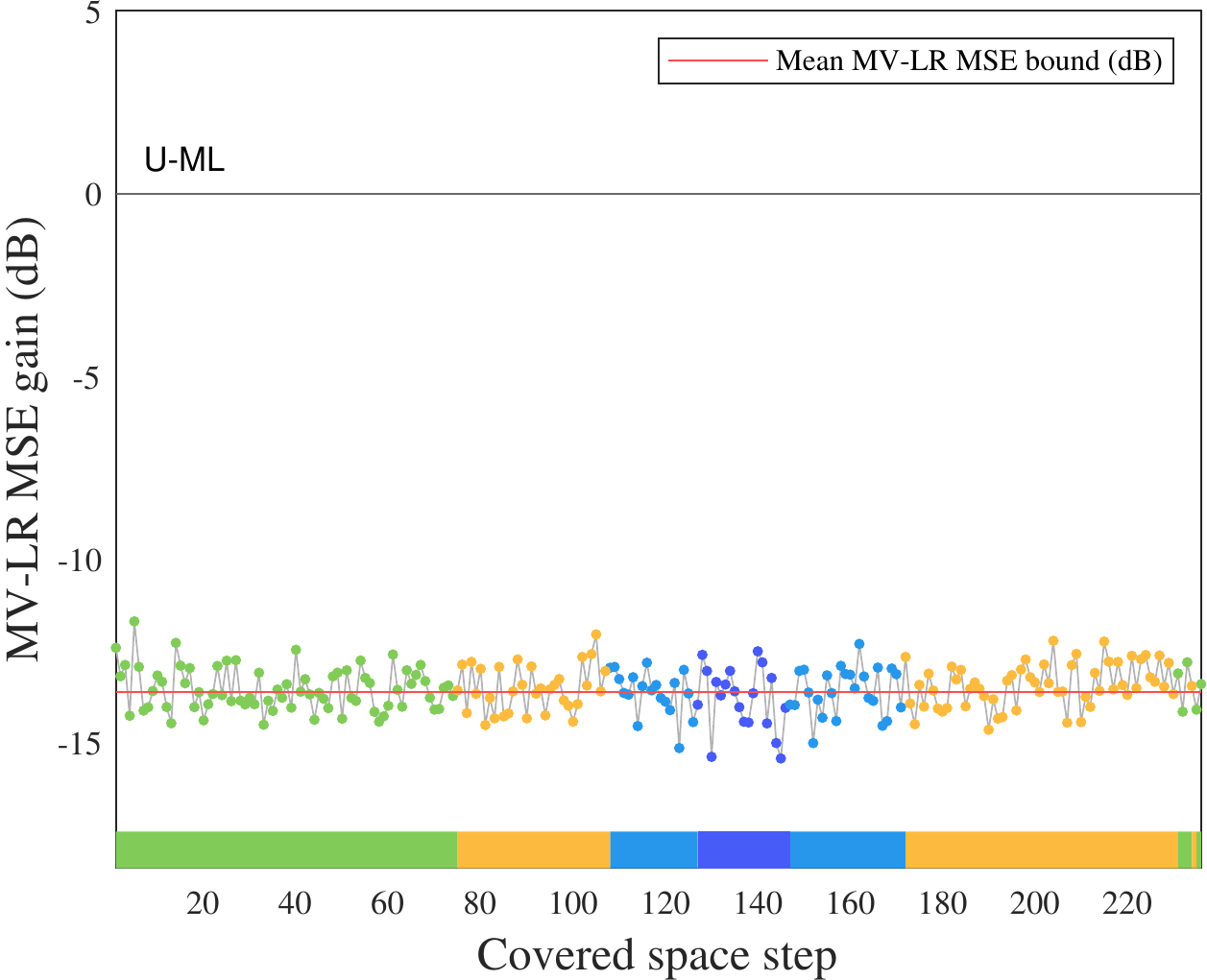}}
    \hfill
    \vrule
    \hfill
    \subfloat[\label{2b}]{%
       \includegraphics[width=0.42\linewidth]{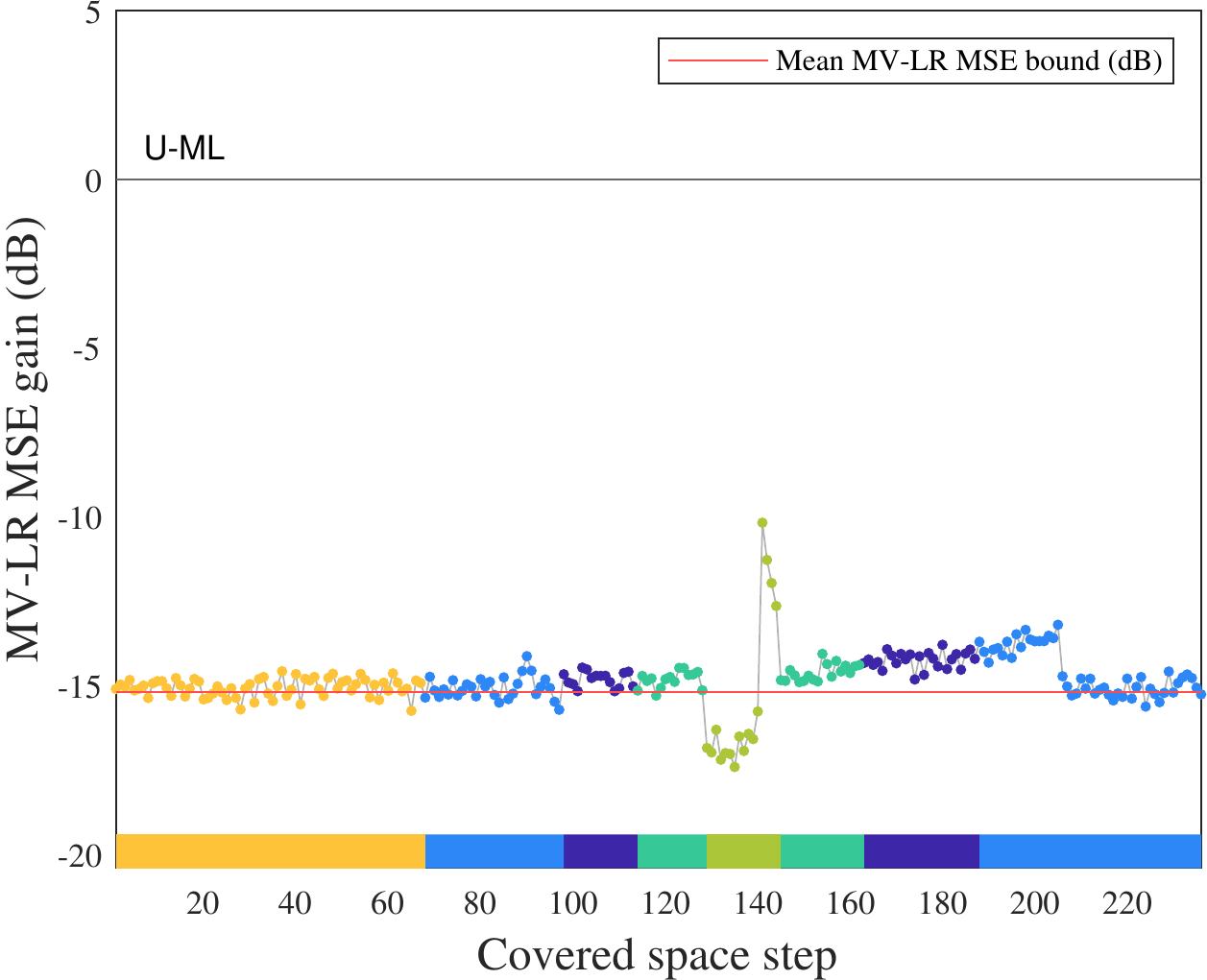}}
       
    \caption{Performance of the proposed algorithm at $1$ MHz channel bandwidth (left) for $K=7$, and $50$ MHz channel bandwidth (right) for $K = 8$: spatial representation of clusters and reference trajectory (a,b); silhouette coefficients of the retrieved clusters (average silhouette coefficient in dashed line) (c,d); MSE performance comparison on a reference trajectory realization (e,f).}
    \label{fig:clustering_on_map_and_MSE_on_trajectory}
\end{figure*}

The recurrent vehicle passages in the cell are generated using the SUMO (Simulation of Urban MObility) software \cite{SUMO2018}, providing position, velocity and heading of vehicles over time for different realistic trajectories, exemplified in Fig. \ref{fig:map_with_trajectories}. The MIMO channel data over the trajectories is generated with the Altair WinProp ray-tracing software \cite{WinProp}. The algorithm has been trained using a dataset of $N=5000$ received training sequences, sampled over the vehicular trajectories at $0$ dB of SNR. The clustering-based MV-LR performance has been evaluated in terms of Mean Squared Error (MSE), defined as:
\begin{equation}\label{eq:MSE}
    \mathrm{MSE} = \frac{\mathbb{E}[\lVert\mathbf{h}_\ell - \widehat{\mathbf{h}}_\ell\rVert^2]}{\mathbb{E}[\lVert\mathbf{h}_\ell\rVert^2]},
\end{equation}
where the channel estimate $\widehat{\mathbf{h}}_\ell$ can be either MV-LR or U-ML. The MV-LR is asymptotically ($L\rightarrow \infty$) lower bounded as detailed in \cite{Cerutti2020}, not reported here for brevity. In the results, the MV-LR MSE bound is averaged over the whole trajectory length.




Fig. \ref{fig:clustering_on_map_and_MSE_on_trajectory} summarizes the results. Figs. \ref{1a} and \ref{2a} represent the extracted clusters---depicted with different colors---over the geographical map for the selected urban scenario, and the tested trajectory (dashed arrow), not comprised in the training dataset.
Using the silhouette method \cite{rousseeuw1987silhouettes}, we chose a suitable number of clusters $K$ such that a sufficient number of relevant data points leads to the convergence of the MV-LR algorithm within each cluster. Then, considering the MV-LR MSE performance, we determined that, for the selected urban scenario, a number of clusters $K=7$ for $1$ MHz channel bandwidth and $K=8$ for $50$ MHz channel bandwidth yield a substantial MSE improvement over U-ML channel estimates.

The related silhouette coefficients are shown in Figs. \ref{1c} and \ref{2c} for each cluster. On the vertical axis, the width of each silhouette is representative of the number of samples assigned to the corresponding cluster, while the vertical dashed line is the average silhouette coefficient. As reported in Section \ref{sec:method}, a clustering that suitably distributes the dataset points among clusters, as here, increases the number of available received training sequences for accurately estimating the MV-LR orthonormal bases. It is worth noticing that, in Figs. \ref{1a} and \ref{2a}, the colored clustered points on the map are not necessarily representative of UEs positions; they depict the invariance regions of the channel estimates in the ST domain for the retrieved clustering, assuming different spatial configurations.



Figs. \ref{1b} and \ref{2b} show the MV-LR MSE performance of the proposed method over the reference trajectory, normalized to the U-ML one. The chosen trajectory is sampled over the covered space by $0.5$ m steps. To compare the achieved performance with the theoretical lower bound, we plot the mean MV-LR MSE bound (red line), averaged over all the trajectory steps. As can be seen, MV-LR outperforms U-ML by achieving $\approx 15$ dB less of MSE, attaining on average the theoretical bound. A similar behavior has been observed for all the other testing trajectories, not reported here, confirming the effectiveness of the proposed clustering-based MV-LR channel estimation method.

\section{Conclusion}

This paper proposes a novel clustering-based MV-LR channel estimation method for 6G V2X. By clustering, through a K-medoids approach, a dataset of received training sequences from multiple UEs, the BS learns, in a completely unsupervised way, to aggregate training sequences sharing similar ST subspaces, to estimate the cluster-specific ST MIMO channel eigenmodes without the knowledge of UEs' geographical positions. For a number of clusters suitably selected by means of the silhouette method, numerical results show remarkable benefits in terms of MSE, with an average reduction of $\approx 15$ dB with respect to the U-ML channel estimates, thus attaining the theoretical LR lower bound. Future investigations will extend the proposed method to hybrid MIMO systems and to propagation affected by blockage.


\section*{Acknowledgements}
The research has been carried out in the framework of Huawei-Politecnico di Milano Joint Research Lab.

\bibliographystyle{IEEEtran}
\bibliography{bibliography}

\end{document}